\documentstyle[11pt,newpasp,twoside,epsf]{article}
\def\NH{$N_{\rm HI}$~} 
\def\kms{km\,s$^{-1}$} 

\def \hi {H\,{\sc i~}} 

\def\edcomment#1{\iffalse\marginpar{\raggedright\sl#1\/}\else\relax\fi}
\reversemarginpar

\begin{document}
\title{CHVCs: Galactic Building Blocks at z = 0}
 \author{Robert Braun}
\affil{Netherlands Foundation for Research in Astronomy, P.O. Box 2,
 7990 AA Dwingeloo, The Netherlands}
\author{W. Butler Burton}
\affil{Sterrewacht Leiden, P.O. Box 9513, 2300 RA Leiden, The Netherlands}

\begin{abstract}
  A distinct sub-class of anomalous velocity \hi emission features has
  emerged from recent high quality surveys of the Local Group
  environment, namely the compact high velocity clouds (CHVCs). A
  program of high-resolution imaging with the Westerbork array and the
  Arecibo telescope has begun to provide many insights into the nature
  of these objects. Elongated core components with a velocity gradient
  consistent with rotation (V$_{Rot}\sim$ 15 \kms) are seen in many
  objects. Comparison of volume and column densities has allowed the
  first distance estimates to be made (600$\pm$300~kpc). The objects
  appear to be strongly dark-matter dominated with dark-to-gas mass
  ratios of 30--50 implied if the typical distance is 700~kpc. 
\end{abstract}

\section{Introduction}

The nature of anomalous velocity \hi discovered during $\lambda$21~cm
surveys of the Galaxy has been the subject of much debate in the past
three decades. There appear to be at least four reasonably distinct
classes of this gas, namely: (1) Localized outflows and subsequent
inflows associated with massive star formation which are commonly
referred to as a ``galactic fountain'' (Shapiro \& Field, 1976;
Bregman, 1980). (2) The stream of tidal debris from the interaction of
the Galaxy with the Magellanic Clouds is another major source of
anomalous velocity \hi\/. Recent imaging (Putman \& Gibson, 1999) has
begun to reveal the true extent of this debris system, and allow it to
be distinguished from other components. (3) The extended HVC complexes
(named A, C, H, M, \dots) which span ten's of degrees on the sky and
have recently been determined to have both nearby distances of about
10~kpc (Van Woerden et al. 1999) and rather low ($\sim$0.1 solar) metal
abundance (Wakker et al. 1999). These diffuse \hi structures appear to
be currently merging with the Galaxy and each account for some 10$^7$
M$\odot$ of fresh gas. And finally (4) the system of compact high
velocity clouds (CHVCs) cataloged by Braun \& Burton (1999).

The CHVCs form a distinct class of compact, high-contrast \hi emission
features at anomalous velocity, which can be readily distinguished from
the more diffuse components in the recent high quality surveys; the
Leiden/Dwingeloo Survey in the North (Hartmann \& Burton 1997) and the
Parkes Multibeam Survey in the South (see Putman \& Gibson 1999). Their
average angular size is only 1~degree FWHM and the total velocity width
30~\kms~ FWHM. The 65 confirmed members of the CHVC class share a
well-defined kinematic pattern. A global search for the reference frame
that minimizes the line-of-sight velocity dispersion of the system,
returns the Local Group Standard of Rest with high confidence. In this
system the velocity dispersion of the population is only 70~\kms,
although it is in-falling toward the Local Group barycenter at about
100~\kms. These properties suggest that the objects are: (1) associated
with the Local Group rather than the Galaxy as such, (2) that they are
likely to reside at quite substantial distances and (3) have as yet had
little tidal interaction with the more massive Local Group members.

\begin{figure}
\plotone{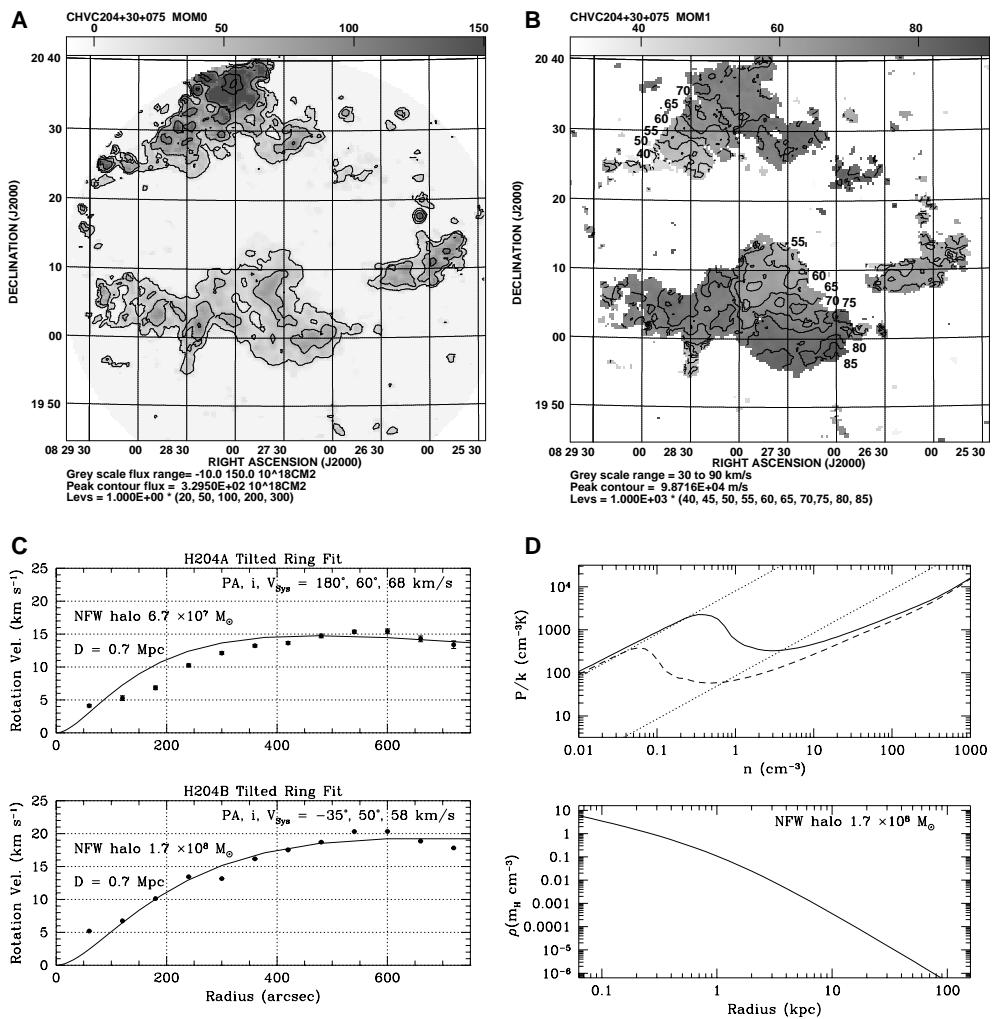}

\caption{{\bf A:}~ Westerbork image of CHVC\,$204\!+\!30\!+\!075$ 
  showing \NH at 1~arcmin
  resolution; contours are drawn at levels of 20, 50, 100,
  200, and $300 \times 10^{18}$~cm$^{-2}$.  {\bf B:}~
  Intensity--weighted line--of--sight velocity, with contours of
  $v_{\rm LSR}$ drawn in steps of 5 \kms~ from 40 to 85 \kms.  {\bf C:}
  Rotation velocities fit to the two principal components of
  CHVC\,$204\!+\!30\!+\!075$.  The solid lines show the rotation curves
  of Navarro, Frenk, \& White (1997) cold--dark--matter halos of the
  indicated mass and 0.7~Mpc distance.  {\bf D: top}~ Equilibrium
  temperature curves for \hi in an intergalactic environment
  characterized by a metallicity of 10\% of the solar value and a
  dust--to--gas ratio of 10\% of that in the solar neighborhood,
  calculated for two values of the neutral shielding column depth,
  namely $10^{19}$~cm$^ {-2}$ (solid line) and $10^{20}$~cm$^{-2}$
  (dashed line).  {\bf D: bottom}~ Mass volume density, in units of
  hydrogen nuclei per cubic centimeter, as a function of radius for an
  NFW cold dark matter halo.  }
\end{figure}

\section{High Resolution Imaging}

A program of high resolution imaging of the CHVCs has begun to provide
further insights into the nature of these sources. An initial
sub-sample of six CHVCs was imaged with the upgraded Westerbork array
(WSRT) in December 1998. A complete description of those results can be
found in Braun \& Burton (2000). The sub-sample was chosen to span a
wide range of positions on the sky and in line-of-sight velocity, as
well as gas content and \hi linewidth.  An additional eight sources
have been imaged with the WSRT during the summer of 1999, specifically
targeting objects suspected of having very bright and narrow \hi
emission features.  A sample of ten CHVCs have also been imaged with
the upgraded Arecibo telescope in November of 1999. In addition to
fully-sampled images of the central square degree, a very deep
cross-cut of two degree length was obtained for each source reaching an
rms sensitivity of about $\Delta$\NH~=~10$^{17}$~cm$^{-2}$ at 10~\kms~
resolution. Full accounts of these recent results will be given
elsewhere.

\subsection{Core/Halo Morphology and Physics}

All of the imaged sources share a number of common properties. Each
source contains between one and ten compact cores (of 1 to 10 arcmin
extent) embedded in a diffuse halo of about 1~degree diameter. An
average of 40\% of the \hi line flux arises in the compact cores
(although the range varies between 1 and 60\%). The peak column density
in these cores varies between 10$^{19}$ and 10$^{21}$~cm$^{-2}$. The
fractional surface area of each source which exceeds a column density
of 5$\times10^{18}$ is only 15\% on average, although this also varies
greatly from source to source. The remainder of the \hi emission arises
in a diffuse halo which reaches peak column densities of
$10^{18.5}$--$10^{19.5}$~cm$^{-2}$ but declines smoothly with an exponential
profile to values of $10^{17.5}$~cm$^{-2}$ or less (as illustrated in Fig.2).

The resolved \hi linewidths toward the core components (with 1~arcmin
resolution) are typically only 5~\kms~ FWHM (but range from as little as
2 to as much as 15~\kms). All of the core linewidths are sufficiently
narrow that it is possible to unambiguously identify these with the
Cool Neutral Medium phase of \hi, with typical kinetic temperature of
100~K. The diffuse halos, on the other hand, all have resolved \hi
linewidths (with 3~arcmin resolution) of 21--25~\kms~ FWHM, as seen in
Fig.2d. The one-to-one correspondence of halo emission with linewidths
equal to or exceeding the thermal linewidth of 8000~K gas (21~\kms~
FWHM) provides an unambiguous identification of these halos with the
Warm Neutral Medium phase of \hi. A shielding column of WNM is one
prerequisite for the condensation of CNM cores in calculations of \hi
thermodynamics (eg. Wolfire et al. 1995). The second prerequisite for
CNM condensation is a sufficiently high thermal pressure, in excess of
about P/k~=~100~cm$^{-3}$K. Both of these aspects are illustrated in
the \hi phase diagram shown in Fig.1d. An adequate shielding column of WNM
is clearly detected in the CHVCs, as just discussed above. We will
return to the question of an adequate thermal pressure below.

Another general result is that the CNM core components are dynamically
decoupled from their WNM halos. This is illustrated in Fig.2, where a
substantial velocity gradient of some 30~\kms~ is seen within the major
core component of CHVC158$-$39$-$285. Such large gradients are not
observed in the halo gas, which instead is always centered near the
systemic velocity of the system. This dichotomy suggests that while the
cores may represent flattened rotating disks, the halos are more nearly
spherical distributions. 

\begin{figure}
\plotone{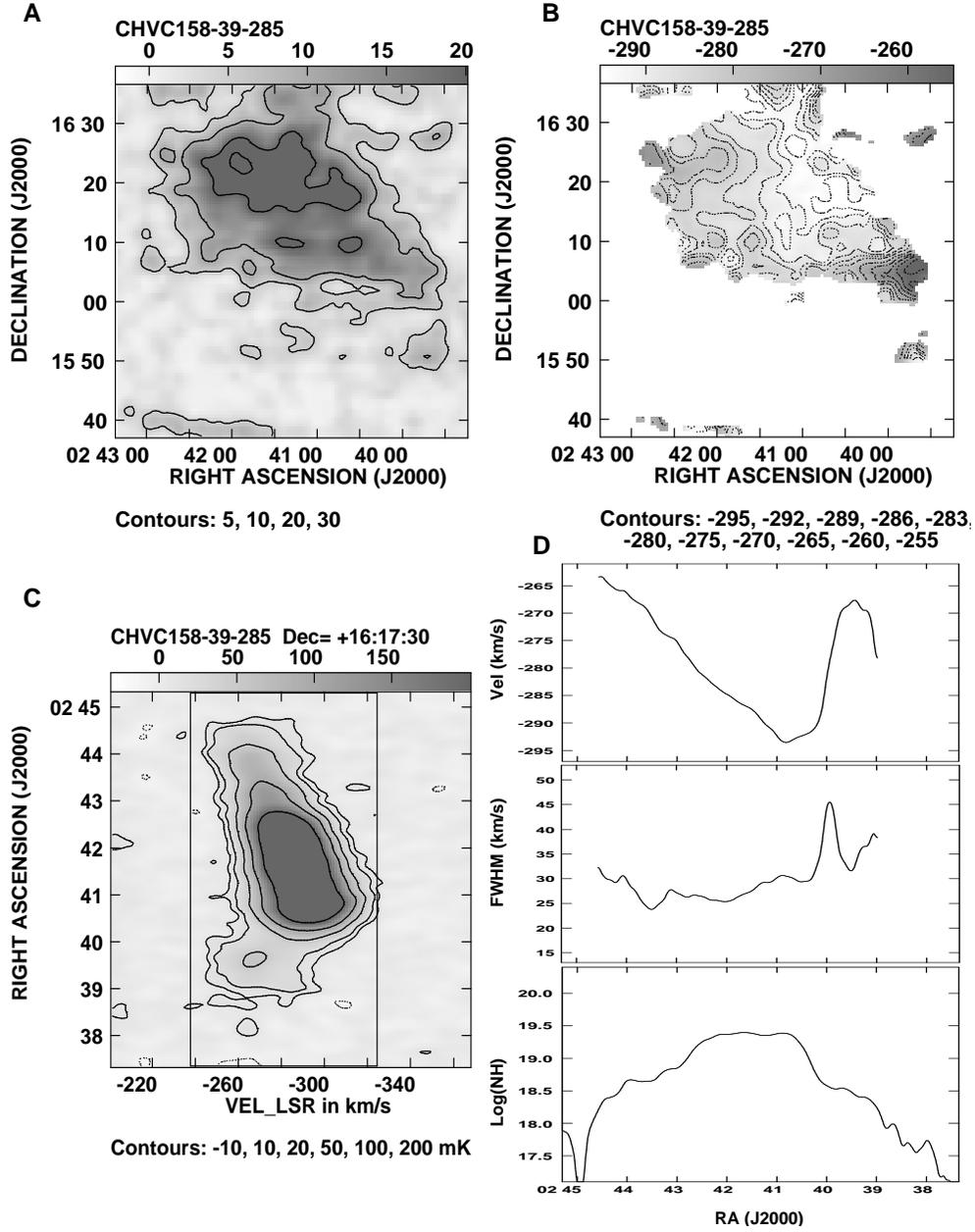}

\caption{{\bf A:} Arecibo image of CHVC\,$158\!-\!39\!-\!285$
  showing \NH at 3 arcmin resolution; contours are drawn at levels of
  5, 10, 20, and $30 \times 10^{18}$~cm$^{-2}$.  {\bf B:}~
  Intensity--weighted line--of--sight velocity, with contours of
  $v_{\rm LSR}$ drawn at the indicated velocities.  {\bf C:} Position,
  velocity cut at a fixed declination of +16:17:30. Contours are in
  units of brightness temperature.  {\bf D:} Variation of line-of-sight
  velocity, velocity FWHM and \NH with position.  }
\end{figure}

\subsection{Distance, Metallicity and Dark Matter Content}

One of the objects imaged with the WSRT, CHVC125+41$-$207, has proven
particularly interesting for several reasons. This object has a number
of opaque core components of only $\theta$~=~90\arcsec angular size, a
brightness temperature in \hi emission of 75~K, and linewidths so
narrow (less than 2~\kms~ FWHM) that the kinetic temperature
(T$_k$~=~85~K), \hi opacity ($\tau$~=~2) and column density
(\NH~=~10$^{21}$cm$^{-2}$), can be accurately derived. A good estimate
of the volume density (n$_{H}$~=~2$\pm$1~cm$^{-3}$) for this object has
been made by modeling the thermodynamics (Wolfire et al., priv. comm.).
With these quantities in hand and only the assumption of crude
spherical symmetry for the cores, it is possible to estimate the source
distance from D~=~\NH/(n$_{H}$$\theta$)~=~600$\pm$300~kpc.

The same object has also allowed the first measurement of CHVC
metallicity, since the bright UV source Mrk~205 is located behind its
diffuse \hi halo. Bowen, Blades and Pettini (1995) detect unsaturated MgII
absorption, with 0.15 \AA~EW at the CHVC velocity, while we detect
a neutral column \NH~=~5$\times$10$^{18}$ cm$^{-2}$ at the same
position. The implied metal abundance is about 0.05 solar.

Although many of the CHVCs have only one or two core components, a
handful of these objects might best be called CHVC clusters. As many as
ten cores are seen in close proximity (within 30--50~arcmin), each with
its own systemic velocity and internal kinematics (which we address
below) but sharing the same diffuse halo of shielding gas. The most
extreme case is that of CHVC115+13$-$275, where the cores have relative
velocities as high as 70 \kms~ and angular separations as large as
30~arcmin. If these are gravitationally bound systems (and their cool
gas content and isolation on the sky make this seem plausible) then
they have a dynamical mass, $M_{\rm dyn}=Rv^2/G=2.3\times 10^5R_{\rm
  kpc}v_{\rm km/s}^2$, and gas mass $M_{\rm gas}=1.4~M_{\rm
  HI}=3.2\times 10^5S~D_{\rm Mpc}^2$. If for example the distance were
0.7~Mpc, then $M_{\rm dyn}=10^{8.93}$ M$_\odot$ and $M_{\rm
  gas}=10^{7.22}$ M$_\odot$. The dark--to--gas mass ratio in this case
has the rather substantial value of $\Gamma$~=~51, which scales with
$1/D$ for other distance estimates.

As already noted above, many of the CHVC cores give an indication for
organized internal kinematics. Many of the cores have a systematic
velocity gradient along the long axis of their elliptical distribution
which is very suggestive of rotation in a flattened disk system.
Several examples are shown in Figs.1 and 2. The best-resolved examples
of this pattern have been subjected to the standard tilted ring fitting
algorithms used in deriving galactic rotation curves. Good solutions
for rotation are found (Fig.1c) which slowly rise over some 6~arcmin to
constant values of 15 to 20~\kms. The shapes of these rotation curves
are in very good agreement with those predicted by Navarro, Frenk \&
White (1997) for a CDM halo at 0.7~Mpc distance (as shown in Fig.1c).
The dark--to--visible mass ratios in these two cases are $\Gamma$~=~36
and 29. 

The presence of a significant dark matter component is important not
only for understanding the kinematics of the individual cores and the
CHVC clusters, but also for understanding the \hi thermodynamics. As
noted earlier, a significant thermal pressure in excess of about
P/k~=~100~cm$^{-3}$K is required to allow condensation of the cool
cores observed in the CHVCs (see Fig.1d, upper). The mass density
provided within the central few kpc of an NFW halo in this mass range
(Fig.1d, lower) is sufficient to provide the required hydrostatic
pressure in combination with the observed warm halo column densities of
about 10$^{19}$cm$^{-2}$ (see also Braun \& Burton 2000).

\acknowledgments  

We are grateful to M.G. Wolfire, A. Sternberg, D. Hollenbach, and C.F.
McKee for providing the equilibrium temperature curves shown in
Fig.1d.  The Westerbork Synthesis Radio Telescope is operated by the
Netherlands Foundation for Research in Astronomy, under contract with
the Netherlands Organization for Scientific Research. The Arecibo
Observatory is part of the National Astronomy and Ionosphere Center,
which is operated by Cornell University under a cooperative agreement
with the National Science Foundation.

\end{document}